\documentclass[prb,twocolumn,showpacs,preprintnumbers,amsmath,amssymb]{revtex4}
\usepackage{graphicx}% Include figure files
\usepackage{dcolumn}% Align table columns on decimal point
\usepackage{bm}% bold math

\begin{document}

%\preprint{revised for PRB ver. 2}

\title{Uncommonly High Upper Critical Field in the Superconducting
KOs$_2$O$_6$ Pyrochlore}%% with Missing Inversion Symmetry }

\author{T. Shibauchi,$^{1}$ L. Krusin-Elbaum,$^2$ Y. Kasahara,$^1$
Y. Shimono,$^1$ Y. Matsuda$^{1,3}$ R.~D. McDonald,$^4$ C.~H.
Mielke,$^4$ S. Yonezawa,$^3$ Z. Hiroi,$^3$ M. Arai,$^5$ T. Kita,$^6$
G. Blatter,$^7$ and M. Sigrist$^7$}

\affiliation{ $^1$Department of Physics, Kyoto University,
Sakyo-ku, Kyoto 606-8502, Japan\\
$^2$IBM T. J. Watson Research Center,
Yorktown Heights, New York 10598, USA\\
$^3$Institute for Solid State Physics, University of Tokyo,
Kashiwa, Chiba 277-8581, Japan\\
$^4$NHMFL, Los Alamos National Laboratory, Los Alamos, New Mexico 87545, USA\\
$^5$National Institute for Materials Sciences,
Namiki 1-1, Tsukuba, Ibaraki 305-0044, Japan\\
$^6$Division of Physics, Hokkaido University, Sapporo 060-0810, Japan\\
$^7$Theoretische Physik, ETH Zurich, CH-8093 Zurich, Switzerland}

%\date{\today}

\begin{abstract}
The entire temperature dependence of the upper critical field
$H_{\rm c2}$ in the $\beta$-pyrochlore KOs$_2$O$_6$ is obtained from
high-field resistivity and magnetic measurements. Both techniques
identically give $H_{\rm c2}(T \simeq 0~{\rm K})$ not only
surprisingly high ($\sim 33$ T), but also the approach to it
unusually temperature-\emph{linear} all the way below $T_{\rm c}$ (=
9.6 K). We show that, while $H_{\rm c2}(0)$ exceeds a simple
spin-singlet paramagnetic limit $H_{\rm P}$, it is well below an
$H_{\rm P}$ enhanced due to the missing spatial inversion symmetry
reported recently in KOs$_2$O$_6$, ensuring that the pair-breaking
here is executed by orbital degrees. {\it Ab initio}
calculations of orbital $H_{\rm c2}$ show that the unusual
temperature dependence is reproduced if dominant $s$-wave 
superconductivity resides on the smaller closed Fermi surfaces.

\end{abstract}

\pacs{74.25.Op, 74.25.Fy, 74.25.Ha, 74.20.Rp}
%74.25.Op   Mixed states, critical fields, and surface sheaths
%74.25.Fy   Transport properties (electric and thermal conductivity, thermoelectric effects, etc.)
%74.25.Ha   Magnetic properties
%74.20.Rp   Pairing symmetries (other than s-wave)

\maketitle

Transition metal oxides, with a nexus of strong electron
correlations and structural diversity in the ways oxygen tetrahedra
and octahedra can be edge- and corner-linked, are well known to host
rather unusual quantum states. High-$T_{\rm c}$ copper oxide
superconductors or manganites are most explored, \cite{Tokura} but
such quantum phenomena are also in evidence in the ``pyrochlore"
structure \cite{Canals} where, in addition, geometrical (spin)
frustration enters in a crucial way. Superconductivity in
$\beta$-pyrochlore oxides $A$Os$_2$O$_6$ discovered not long ago,
with relatively high transition temperatures $T_{\rm c}$ (3.3 K, 6.3
K, and 9.6 K for $A$ = Cs, \cite{YonezawaCs} Rb, 
\cite{YonezawaRb,Kazakov} and K, \cite{YonezawaK} respectively) and
distinctly odd behaviors in the normal state, suggests new physics,
perhaps explicitly connected to this structure.

KOs$_2$O$_6$, with the highest $T_{\rm c}$, appears to be more
unusual than the rest. The resistivity in the normal state has a
pronounced \emph{convex} temperature dependence down to $T_{\rm c}$,
\cite{Hiroi} indicating that electron-phonon scattering is strong
-- likely owing to the rattling motion of `caged' K ions.
\cite{Kunes} The specific heat has a jump $\Delta C/T_{\rm c} =
185$~mJ K$^{-2}$ mol$^{-1}$ at $T_{\rm c}$, but also another (jump)
anomaly at a lower temperature $T_{\rm p} \sim 7.5$ K that has been
attributed to freezing of the K rattle.
\cite{Hiroi_p,Kasahara,Batlogg} In addition, strong electron
correlations show up in important ways in transport and
thermodynamic properties: for example, the thermal conductivity of
KOs$_2$O$_6$ is enhanced in the superconducting state
\cite{Kasahara} (reminiscent of high-$T_{\rm c}$ cuprates), and the
Sommerfeld coefficient $\gamma$ is also largely enhanced
\cite{Hiroi_p,Batlogg} from the band calculation value. \cite{Kunes}

The coexistence of strong electron correlations that prefer an
anisotropic order parameter and strong electron-phonon coupling that
favors a fully gapped $s$-wave ground state may render the workings
of superconducting pairing in KOs$_2$O$_6$ rather uncommon.
Experimentally, the situation appears contradictory: $\mu$SR
measurements \cite{Koda} suggest anisotropic gap
functions with nodes, in sharp contrast to the nodeless gap in
RbOs$_2$O$_6$, \cite{Magishi} while low-temperature thermal
conductivity \cite{Kasahara} -- based on its magnetic field
insensitivity -- is consistent with a fully gapped state.

Indeed, there has been much speculation about possible
(unconventional) modes of pair-breaking in KOs$_2$O$_6$ at low
temperatures. Based on \emph{extrapolated} (from low fields)
estimates of upper critical field $H_{\rm c2}$ in the $T \rightarrow
0$~K limit, suggestions have been made, \cite{Hiroi,Hiroi_p,
Batlogg,Schuck} that the spin contribution to the pair-breaking
must be significant, that $H_{\rm c2}(0)$ in KOs$_2$O$_6$ may exceed
the Pauli paramagnetic limit expected in a spin-singlet
superconductor, that a quantum critical state may enter, and that a
state with a spatially modulated order parameter
(Fulde-Ferrell-Larkin-Ovchinnikov (FFLO) state \cite{FFLO})  may
appear at low $T$ and high magnetic fields.

Here we show that in this pyrochlore system, \emph{missing spatial
inversion symmetry} can uniquely control the pair-breaking process,
leading to unconventional behavior of the upper critical field
without an unconventional pairing mechanism found in some heavy
fermion systems (e.g. CePt$_3$Si (Ref.~\onlinecite{Bauer})) that also lack
inversion symmetry.

We have experimentally reached the low-$T$ high-field limit to
obtain the full temperature dependence of $H_{\rm c2}$ in
KOs$_2$O$_6$. We find that $H_{\rm c2}$ in the $T \rightarrow 0$~K
limit is not only surprisingly high, but also the approach to it
does not display the typical flattening at low $T$. Both high-field
resistivity and magnetic penetration measurements gave us an
identical $H_{\rm c2}(T)$ growing \emph{linearly} with temperature
and reaching $\sim 32$~T in the subKelvin range. This value is
clearly beyond the simple Clogston paramagnetic limit of $H_{\rm
P}\sim 18$~T. \cite{Clogston} Following a remarkable recent
structural finding of broken symmetry in KOs$_2$O$_6$, \cite{Schuck}
we show by relying on experimental inputs how this limit can be
hugely enhanced (up to $\sim 54$~T). This enhancement leaves orbital
pair-breaking protected from spin effects up to very high fields,
with the observed $T$-linear $H_{\rm c2}$ fully consistent with %{\it
%ab initio} calculations of
the orbital contributions from the closed Fermi surfaces.

In this study, we used a block containing several single crystals of
cubic KOs$_2$O$_6$ grown by the technique described in 
Ref.~\onlinecite{Hiroi}. The
resistivity was recorded using a 100~kHz lock-in technique in a 65~T
maximum field, 60~ms pulsed magnet \cite{Krusin} at the National
High Magnetic Field Laboratory (NHMFL in Los Alamos). The magnetic
penetration was measured by a tunnel diode oscillator (TDO)
operating at $f\sim 55$~MHz in an LC tank circuit. \cite{Mielke} A
heterodyne technique was used to beat down the frequency to the
hundreds of kHz range, with the waveform recorded during the pulse.
The sample was inserted in one coil of the pair wound in a
gradiometer configuration [sketched in Fig.~2], and the inductance
change due to the change in the penetration depth was detected by
the shift of the resonance frequency $\Delta f$.

%%%%%%%%%%%%%%%%%%%%%%%%%%%%%%%%%%%FIG 1%%%%%%%%%%%%%%%%%%%%
\begin{figure}%[td]
%\includegraphics[width=87mm]{KOs2O6_Fig1r}%
%\hspace{-10mm}
\includegraphics[width=90mm]{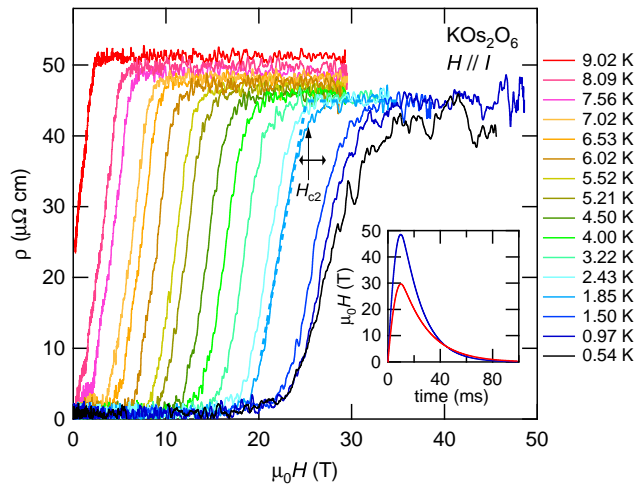}%
\caption{(color online). Field dependence of the resistivity in
KOs$_2$O$_6$. In the experiments, a block of small single crystals
was cut into two pieces: one was used for resistivity and the other
for magnetic measurements, as those in Fig.~2. Inset: Typical
field-pulse profiles used. Magnetic fields up to $\sim 50$~T were
applied in the current direction.}
\end{figure}
%%%%%%%%%%%%%%%%%%%%%%%%%%%%%%%%%%%FIG 1%%%%%%%%%%%%%%%%%%%%

%%%%%%%%%%%%%%%%%%%%%%%%%%%%%%%%%%%FIG 2%%%%%%%%%%%%%%%%%%%%
\begin{figure}%[td]
\includegraphics[width=90mm]{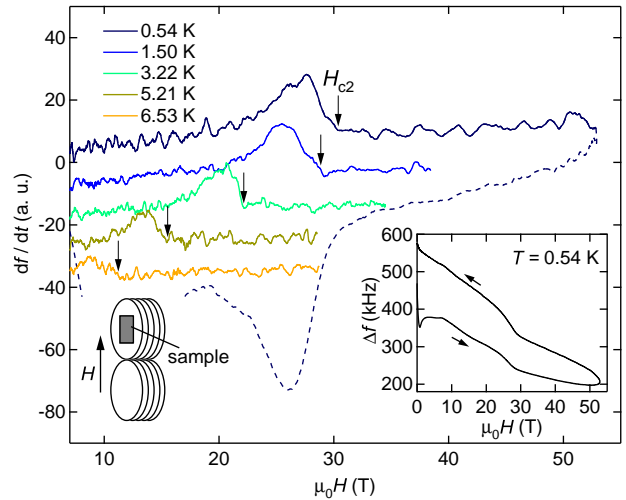}%
\caption{(color online). Time derivative $\textrm{d}f/\textrm{d}t$
of the frequency shift of the tunnel diode oscillator containing the
KOs$_2$O$_6$ sample. The full up- (dashed line) and down-field cycle
is shown for $T=0.54$~K, for which the frequency shift $\Delta f$
vs. field is also displayed in the inset. At other temperatures
(color coded as in Fig.~1), only down-field sweeps are shown and
each curve is vertically shifted for
clarity.%used for determination of the upper critical field.
}
\end{figure}
%%%%%%%%%%%%%%%%%%%%%%%%%%%%%%%%%%%FIG 2%%%%%%%%%%%%%%%%%%%%

Figure~1 shows the field dependence of the resistivity $\rho$ in
KOs$_2$O$_6$. Upon the field sweep, $\rho(H)$ evolves, as expected,
from the superconducting (vortex) state at low fields to the
high-field normal state. The resistive transition is relatively
sharp; a certain amount of broadening is expected
\cite{Blatter,Morozov} since the crystals in a block are weakly
connected. The upper critical field $H_{\rm c2}$ can therefore be
determined in the usual way from the field at which $\rho(H)$ is
fully restored to its normal-state value. In order to remove any
ambiguity in the resistively determined $H_{\rm c2}$ 
(Ref.~\onlinecite{Morozov})
we have further corroborated our results by magnetic measurements.
The inset of Fig.~2 displays the frequency shift $\Delta f$ of the
TDO as a function of $H$. The observed hysteresis is related to the
asymmetry of the field pulse shown in the inset of Fig.~1 -- the
field pulse rise time (10~ms) is much shorter than the fall time
(50~ms). The field direction is perpendicular to the axis of the
coils; in our setup one coil detects the change in the sample and
the other is for the cancellation of the voltage signal from ${\rm
d}B/{\rm d}t$. To bypass the somewhat imperfect cancellation, we plot
${\rm d}f/{\rm d}t$ as a function of field in the main panel of
Fig.~2, where the anomaly (peak) due to the change in the
penetration depth is clearly articulated. The high-field end point
of the peak in ${\rm d}f/{\rm d}t$ corresponds to the field where
the whole sample becomes normal, and hence it is the value of the
upper critical field. The $H_{\rm c2}$ values determined from the up
and down sweeps of the field pulse coincide with each other, which
is a solid indication that our measurements are free from
eddy-current heating $\propto ({\rm d}B/{\rm d}t)^2$.

Our independent resistive and magnetic measurements define a unique
upper critical field line $H_{\rm c2}(T)$ (Fig.~3) which is also
consistent with previous low-field data;
\cite{Hiroi,Hiroi_p,Batlogg,Schuck} we surmise then this
temperature dependence is intrinsic to KOs$_2$O$_6$. $H_{\rm c2}(T)$
has two salient features: (i) its temperature dependence is linear
in $T$, without any visible saturation at low temperatures, and (ii)
it reaches 32~T at the lowest temperature measured (0.5~K) and
unambiguously extrapolates to 33~T in the zero temperature limit,
which corresponds to the coherence length $\xi(0)= 3.2$~nm. To
understand the pair-breaking mode, both of these features need to be
accounted for.

%%%%%%%%%%%%%%%%%%%%%%%%%%%%%%%%%%%FIG 3%%%%%%%%%%%%%%%%%%%%
\begin{figure}%[td]
%\vspace{-10mm} \hspace{-12mm}
\includegraphics[width=90mm]{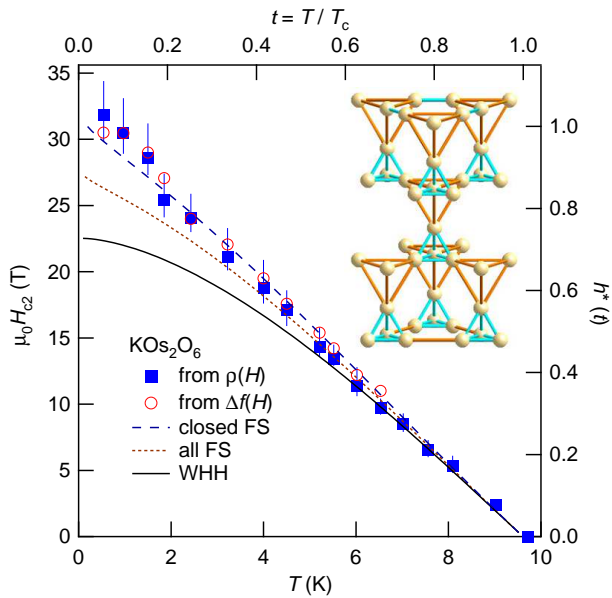}%
%\vspace{5mm} 
\caption{(color online). Temperature dependence of the
upper critical field determined by the resistivity (closed squares)
and magnetic (open circles) measurements. The solid line is the
expected orbital $h^*(t)$ from the conventional WHH formulation. The
dotted and dashed lines are {\it ab initio} calculations for the
maximum $h^*(t)$ ([111] direction) by fitting the initial slope at
$T_c$ and taking into account all the Fermi surfaces (FS) and the
closed FS only, respectively. Inset: Schematic Os network in the
non-centrosymmetric ($F\bar{4}3m$) structure.}
\end{figure}
%%%%%%%%%%%%%%%%%%%%%%%%%%%%%%%%%%%FIG 3%%%%%%%%%%%%%%%%%%%%

Since this upper limiting field is so large, let us attempt a more
realistic estimate of the Pauli paramagnetic limiting field $H_{\rm
P}$. At $H_{\rm P}$, Cooper pairs are broken apart by the Zeeman
splitting produced by the magnetic-field coupling to the electron
spins. This takes place when the Zeeman energy reaches the
condensation energy $U_{\rm c}=N(0)\Delta^2/2 = H_{\rm c}^2/8\pi $
[$N(0)$ is the density of states at the Fermi level, $\Delta$ is the
superconducting gap, and $H_{\rm c}$ is the thermodynamic critical
field]:
\begin{equation}
U_{\rm c} = [\chi_{\rm n}-\chi_{\rm s}(T)]H_{\rm P}^2/2.
\end{equation}
Here $\chi_{\rm n}=g^2\mu_{\rm B}^2N(0)/2$ is the Pauli spin
susceptibility in the normal state ($\mu_{\rm B}$ is Bohr magneton)
and $\chi_{\rm s}(T)$ is the spin susceptibility in the
superconducting state. In spin-singlet superconductors, $\chi_{\rm
s}(T)$ follows the Yoshida function which vanishes at $T=0$~K.
Simple calculations within the weak-coupling BCS theory with
$\Delta=1.76k_{\rm B}T_{\rm c}$ and the assumption $g=2$ give the
well-known result \cite{Clogston} $H_{\rm P}~[{\rm in~ Tesla}]= 1.85
T_{\rm c}~[{\rm in~ Kelvin}]$. In KOs$_2$O$_6$, this limit is 17.8
T, clearly much lower than the observed $H_{\rm c2}(0)$.

We may improve on this estimate by making use of experimental
parameters \cite{Zuo} for the susceptibility $\chi_{\rm n}$ and in
the determination of the condensation energy $U_{\rm c}$: the
normal-state Pauli susceptibility $\chi_{\rm n} \approx 4.2\times
10^{-4}$~emu/mol has been measured \cite{Batlogg,Hiroi_p} just above
$T_{\rm c}$. 
The specific heat jump $\Delta C/T_{\rm c} = 185$~mJ 
K$^{-2}$ mol$^{-1} = [d H_{\rm c}/dT]_{T_{\rm c}}^2/4\pi$ at $T_{\rm
c}$ (Ref.~\onlinecite{Hiroi_p,deGennes}) gives an estimate for the $T=0$
thermodynamic critical field $H_{\rm c}(0) \approx 0.26$~T. Using
Eq.~(1), this results in a larger value $H_{\rm P} \approx 31$~T, 
\cite{Wilson}
somewhat higher than the estimate $H_{\rm P} \approx 27$~T by
Br\"uhwiler {\it et al.} \cite{Batlogg} using strong coupling
corrections. \cite{Orlando} These values are near but still below
the experimental $H_{\rm c2}(0) \approx 33$~T. We point out that in
usual spin-singlet superconductors $H_{\rm c2}(T)$ tends to saturate
below $H_{\rm P}$, \cite{CeCoIn5} which appears to contradict our data.

At first glance, this would suggest spin-triplet superconductivity
for which $\chi_{\rm s}(T)$ remains of order of the normal-state
value, pushing the Pauli limit towards higher fields. Rather than
invoking unconventional pairing, an alternative scenario providing a
finite  $\chi_{\rm s}(T=0)$ derives from the recently reported
non-centrosymmetric crystal structure of KOs$_2$O$_6$ by Schuck
\textit{et al}.;\cite{Schuck} they found a volume deviation from
an ideal $\beta$-pyrochlore lattice in Os tetrahedral and O
octahedral networks and found the structure to be cubic with
$F\bar{4}3m$ space group.

The lack of inversion symmetry [visualized by the Os network in the
inset of Fig.~3] affects the electronic properties through the
appearance of an {\em antisymmetric spin-orbit coupling} (ASOC) term
$\alpha\sum_{\vec{k},s,s'} \vec{g}(\vec{k})\vec{\sigma}_{ss'}
c^\dag_{\vec{k}s}c_{\vec{k}s'}$ in the Hamiltonian, where $\alpha$
denotes the spin-orbit coupling strength, $\vec{\sigma}$ is the
Pauli matrices vector, $c^\dag_{\vec{k}s}$ ($c_{\vec{k}s}$) creates
(annihilates) an electron with momentum $\vec{k}$ and spin $s$, and
$\vec{g}(\vec{k})$ is a dimensionless vector with
$\vec{g}(-\vec{k})=-\vec{g}(\vec{k})$. Such a term will admix
spin-singlet and spin-triplet pairing \cite{Yuan} and hence modify
the spin susceptibility $\chi_{\rm s}(T)$ in the superconducting
state. \cite{FrigeriAll}

The effect has been extensively studied for the non-centrosymmetric
superconductor CePt$_3$Si; \cite{Bauer} there the susceptibility
$\chi_{\rm s}(T)$ of the spin-singlet state assumes the form of a
spin-triplet material with the $\vec d (\vec{k})$-vector of the
triplet order-parameter replaced by the spin-orbit coupling vector
$\vec{g}(\vec{k})$. \cite{FrigeriAll} With a simple $s$-wave
superconductivity found in the sister compounds RbOs$_2$O$_6$ and
CsOs$_2$O$_6$, it appears natural to start from an $s$-wave scenario
also in the present case. Given the $F\bar{4}3m$ symmetry in
KOs$_2$O$_6$ (as in zincblende), the spin-orbit coupling vector
$\vec{g}(\vec{k})$ has a form: \cite{FrigeriAll} 
\begin{equation}
\vec{g}(\vec{k}) =
[k_x(k_y^2-k_z^2),k_y(k_z^2-k_x^2),k_z(k_x^2-k_y^2)]/k_{\rm F}^3, 
\end{equation}
with $k_{\rm F}$ the Fermi wave vector.

In KOs$_2$O$_6$, we expect a fairly large $\alpha$ from the heavy Os
atoms, allowing us to use the spin-triplet state expression in the
determination of $\chi_{\rm s}(0)$, with the replacement 
$\vec d (\vec{k}) \rightarrow \vec{g}(\vec{k})$ as noted above. 
Following the calculations formulated in \cite{FrigeriAll} with 
$\vec{g}(\vec{k})$ in Eq.~(2), we obtain the value
$\chi_{\rm s}(0)= (2/3) \chi_{\rm n}$. The right hand side of our
Eq.\ (1) then is reduced by a factor 1/3, resulting in a
paramagnetic limiting field $H_{\rm P}$ enhanced by a factor of
$\sqrt{3}$. Taking our above estimate of $31$ T based on
experimental values for $U_{\rm c}$ and $\chi_{\rm n}$, we find an
enhanced limiting field $H_{\rm P}\sim 54$~T, way beyond the
observed value of $H_{\rm c2}(0)$. \cite{Wilson}
This large $H_{\rm P}$ then
resides sufficiently far above the measured value $H_{\rm c2}(0)
\simeq 33$ T and thus protects the orbital upper critical field
$H_{\rm c2}(T)$ from spin effects.

The remaining question is how the orbital effects can enforce the
observed linear temperature dependence. The orbital
depairing is usually well described by the
Werthamer-Helfand-Hohenberg (WHH) theory, \cite{WHH} where the
reduced critical field
%\begin{equation}
$h^*(t)=\frac{H_{\rm c2}(t)} {-{\rm d}H_{\rm c2}(t)/{\rm d}t|_{t=1}}$
%\end{equation}
saturates to $h^*(0)=0.727$ in the clean limit. This $h^*(t=T/T_{\rm
c})$, plotted as solid line in Fig.~3, clearly deviates from the
experimental data at low $T$.

Recently Kita and Arai provided a theoretical framework that allows
for {\it ab initio} calculations of orbital $H_{\rm c2}$ accounting
for electronic band-structure effects. \cite{Kita} Band structure
calculations for the $A$Os$_2$O$_6$ compounds \cite{Kunes} unveil
two kinds of Fermi surfaces (FS): one is the connected FS with necks
along the three-fold axis, while the other involves closed sheets
centered on the $\Gamma$ point. Taking these Fermi surface shapes
into account, we performed {\it ab initio} calculations of $h^*(t)$.
We find that the FS anisotropy gives a maximum $H_{\rm c2}$ in the
[111] direction, which we compare with the experimental $H_{\rm c2}$
taken as the field where the \emph{whole} sample becomes normal.
\cite{Hiroi}

Calculated results including all Fermi surfaces (connected and
closed) are shown in Fig.~3. This $h^*(t)$ still deviates from the
$H_{\rm c2}(T)$ data at low temperatures. In contrast, if we ignore
the connected surface and calculate $h^*(t)$ for the closed surfaces
alone, we obtain an essentially $T$-linear $h^*(t)$ without
saturation at low $T$, in very good agreement with the data. We
conclude then, that depairing at the upper limiting field is
enforced by the orbital degrees of freedom, without interferences
from the Pauli limit, and that the superconducting pairing mainly
occurs on the closed Fermi surfaces.

Our finding of orbitally limited $H_{\rm c2}$ is compatible with the
fully gapped superconductivity suggested by thermal conductivity
measurements. \cite{Kasahara} In the spin-triplet case, a nodeless
gap is possible for the $\vec{d}=(k_x, k_y, k_z)=\vec{d}_{\rm BW}$
(known as Balian-Werthamer state). This state, however, is easily
suppressed by the ASOC term that satisfies
$\vec{g}(\vec{k})\cdot\vec{d}_{\rm BW}=0$. \cite{footnote} This
strongly suggests that the $s$-wave spin-singlet component is
dominant in KOs$_2$O$_6$, which is likely mediated by the strong
electron-phonon coupling that wins over the electron correlations.

Lastly we comment on new vortex phases that can arise. In
CePt$_3$Si, the ASOC forms a helical vortex phase, \cite{Kaur}
analogous to the FFLO state with a finite net momentum of Cooper
pairs. \cite{FFLO} Thus, there is an intriguing expectation that a
new vortex state may also appear in KOs$_2$O$_6$. \cite{FrigeriAll}
So in summary, our results highlight a profound influence of broken
spatial inversion symmetry on the nature of
%superconducting
pair-breaking in KOs$_2$O$_6$.

%Acknowledgments
We acknowledge fruitful discussions with S. Fujimoto, M. Takigawa, 
Y. Yanase, P.A. Frigeri, B. Batlogg, M. Br\"uhwiler, J. Karpinski and K.
Rogacki. This work was partly supported by Grants-in-Aid for
Scientific Research from MEXT and by the Swiss National Fonds,
including the NCCR MaNEP. After completion of this work, we became
aware of high-field penetration depth data by Ohmichi {\it et al.}
\cite{Ohmichi} corresponding to similarly high $H_{c2}$.


\begin{references}

%\item[*] Email address: shibauchi@scphys.kyoto-u.ac.jp

\bibitem{Tokura}
Y. Tokura and N. Nagaosa, Science {\bf 288}, 462 (2000).

\bibitem{Canals}
See, for example, B. Canals and C. Lacroix, Phys. Rev. B {\bf 61},
1149 (2000).

\bibitem{YonezawaCs}
S. Yonezawa, Y. Muraoka, and Z. Hiroi, J. Phys. Soc. Jpn. {\bf 73},
1655 (2004).

\bibitem{YonezawaRb}
S. Yonezawa, Y. Muraoka, Y. Matsushita, and Z. Hiroi, J. Phys. Soc.
Jpn. {\bf 73}, 819 (2004).

\bibitem{Kazakov}
S. M. Kazakov \emph{et al.},
%N. D. Zhigadlo, M. Br\"uhwiler, B. Batlogg, and  J. Karpinski,
Supercond. Sci. Technol. {\bf 17}, 1169 (2004).

\bibitem{YonezawaK}
S. Yonezawa \emph{et al.},
%Y. Muraoka, Y. Matsushita, and Z. Hiroi,
J. Phys. Condens. Matter {\bf 16}, L9 (2004).

\bibitem{Hiroi}
Z. Hiroi \emph{et al.},
%S. Yonezawa, J. Yamaura, T. Muramatsu, and Y. Muraoka,
J. Phys. Soc. Jpn. {\bf 74}, 1682 (2005);
%Z. Hiroi, S. Yonezawa, J. Yamaura, T. Muramatsu, 
%Y. Matsushita, and Y. Muraoka,
J. Phys. Soc. Jpn. {\bf 74}, 3400 (2005). The weakly anisotropic
crystallites in the sample block are most likely randomly aligned.
In our study we use a similar sample block, so we experimentally
define $H_{c2}$ as the maximum field where the superconducting
current has completely vanished.

\bibitem{Kunes}
J. Kune$\check{\rm s}$, T. Jeong, and W. E. Pickett, Phys. Rev. B
{\bf 70}, 174510 (2004).

\bibitem{Hiroi_p}
Z. Hiroi {\it et al.},
% S. Yonezawa, Y. Nagao, and J. Yamaura, 
preprint.

\bibitem{Batlogg}
M. Br\"uhwiler, S. M. Kazakov, J. Karpinski, and B. Batlogg,
Phys. Rev. B {\bf 73}, 094518 (2006). 

\bibitem{Kasahara}
Y. Kasahara {\it et al.}, % Y. Shimono, T. Shibauchi, Y. Matsuda, S.
%Yonezawa, Y. Muraoka, and Z. Hiroi,
%cond-mat/0603036.
Phys. Rev. Lett. {\bf 96}, 247004 (2006).

\bibitem{Koda}
A. Koda {\it et al.},
%W. Higemoto, K. Ohishi, S. R. Saha, R. Kadono, S. Yonezawa, Y.
%Muraoka, and Z. Hiroi,
J. Phys. Soc. Jpn {\bf 74}, 1678 (2005).

%\bibitem{Arai}
%K. Arai {\it et al.},
%J. Kikuchi, K. Kodama. M. Takigawa, S. Yonezawa, Y.Muraoka, and Z.
%Hiroi,
%cond-mat/0411460. % (unpublished).

\bibitem{Magishi}
K. Magishi {\it et al.},
%J. L. Gavilano, B. Pedrini, J. Hinderer, M. Weller,
% H. R. Ott, S. M. Kazkov, and J. Karpinski,
Phys. Rev. B {\bf 71}, 024524 (2005).

\bibitem{Schuck}
G. Schuck {\it et al.},
%S. M. Kazakov, K. Rogacki, N. D. Zhigadlo, and J. Karpinski,
Phys. Rev. B {\bf 73}, 144506 (2006).

\bibitem{FFLO}
P. Fulde, and R. A. Ferrel, Phys. Rev. {\bf 135A}, 550 (1964); A. I.
Larkin, and Y. N. Ovchinnikov, Sov. Phys. JETP {\bf 20}, 762 (1965).

\bibitem{Bauer}
E. Bauer {\it et al.}, %G. Hilscher, N. Michor, Ch. Paul, E. W.
%Scheidt, A. Gribanov, Yu. Seropegin, H. Noeel, M. Sigrist, and P.
%Rogl,
Phys. Rev. Lett. {\bf 92}, 027003 (2004). CePt$_3$Si has $P4mm$
space group and a Rashba-type coupling $\vec{g}(\vec{k})\propto
(k_y, -k_x, 0)$.

\bibitem{Clogston}
A. M. Clogston, Phys. Rev. Lett. {\bf 9}, 266 (1962).
%; B. S. Shandrasekhar, Appl. Phys. Lett. {\bf 1}, 7 (1962).

\bibitem{Krusin}
L. Krusin-Elbaum, T. Shibauchi, and C.~H. Mielke, Phys. Rev. Lett.
{\bf 92}, 097005 (2004).

\bibitem{Mielke}
C. H. Mielke {\it et al.},
%J. Singleton, M.-S. Nam, N. Harrison, C. C. Agosta, B. Fravel, and
%L. K. Montgomery,
J. Phys. Condens. Mat. {\bf 13}, 8325 (2001).

\bibitem{Blatter}
G. Blatter {\it et al.},
%M. V. Feigel'man, V. B. Geshkenbein, A. I. Larkin, V. M. Vinokur,
Rev. Mod. Phys. {\bf 66}, 1125 (1994).

\bibitem{Morozov}
N. Morozov {\it et al.},
%L. Krusin-Elbaum, T. Shibauchi, L. N. Bulaevskii, M. P. Maley, Yu.
%I. Latyshev, and T. Yamashita,
Phys. Rev. Lett. {\bf 84}, 1784 (2000).

\bibitem{Zuo}
F. Zuo {\it et al.},
%J. S. Brooks, R. H. McKenzie, J. A. Schlueter, and J. M. Williams,
Phys. Rev. B {\bf 61}, 750 (2000).

\bibitem{deGennes} P. G. de Gennes, in {\it Superconductivity
of Metals and Alloys} (Addison Wesley, New York, 1989), p.\ 18.

\bibitem{Wilson} 
Here, in evaluating $\chi_{\rm n}$ a possible orbital susceptibility
is ignored and consequently the Wilson ratio $R_{\rm W}$ close to 1
is obtained. \cite{Batlogg} Taking $R_{\rm W}\sim 2$, as widely
observed for strongly correlated materials, the $H_{\rm P}$
estimates are reduced only by a factor of $\sim \sqrt{2}$, not
affecting our conclusions. 

\bibitem{Orlando}
T. P. Orlando {\it et al.},
%E. J. McNiff, Jr., S. Foner, and M. R. Beasley,
Phys. Rev. B {\bf 19}, 4545 (1979).

\bibitem{CeCoIn5} T. Tayama {\it et al.},
%A. Harita, T. Sakakibara, Y. Haga, H. Shishido, R. Settai, and Y. Onuki,
Phys. Rev. B  {\bf 65}, 180504(R) (2002). In the heavy-fermion
CeCoIn$_5$ superconductor, the Pauli-limited $H_{c2}(T)$ is markedly
flatter relative to WHH. \cite{WHH}

\bibitem{Yuan}
P. A. Frigeri, D. F. Agterberg, and M. Sigrist, cond-mat/0505108; 
H.~Q. Yuan {\it et al.}, % D. F. Agterberg, N. Hayashi, P. Badica, D.
%Vandervelde, K. Togano, M. Sigrist, and M. B. Salamon,
Phys. Rev. Lett. {\bf 97} 017006 (2006).
%cond-mat/0512601.

\bibitem{FrigeriAll}P. A. Frigeri {\it et al.},
%, D. F. Agterberg, A. Koga, and M. Sigrist,
Phys. Rev. Lett. {\bf 92}, 097001 (2004); P. A. Frigeri {\it et
al.}, %, D. F. Agterberg, and M. Sigrist,
New J. Phys. {\bf 6}, 115 (2004); P.A. Frigeri, doctoral thesis
(2005).

\bibitem{WHH}
%E. Helfand, and N. R. Werthamer, Phys. Rev. {\bf 147}, 288 (1966);
N. R. Werthamer, E. Helfand, P. C. Hohenberg, Phys. Rev. {\bf 147},
295 (1966).

\bibitem{Kita}
T. Kita and M. Arai, Phys. Rev. B {\bf 70}, 224522 (2004).

\bibitem{footnote} The anti-symmetric spin-orbit coupling
suppresses odd pairing states with $\vec d$-vectors not parallel to
$\vec g$, see Ref.~\onlinecite{FrigeriAll}.

\bibitem{Kaur}
R. P. Kaur, D. F. Agterberg, and M. Sigrist, Phys. Rev. Lett. {\bf
94}, 137002 (2005).

\bibitem{Ohmichi}
E. Ohmichi {\it et al.}, J. Phys. Soc. Jpn. {\bf 75}, 045002 (2006).

\end{references}
\end{document}